\documentstyle[12pt]{article}

\advance\hoffset by -7mm  

\makeatletter
\@addtoreset{equation}{section}
\makeatother

\renewcommand{\title}[1]{\null\vspace{25mm}

\noindent{\Large{\bf #1}}\vspace{10mm}

\noindent {\large By }}
\newcommand{\authors}[1]{\noindent{\large #1}\vspace{3mm}

}
\newcommand{\address}[1]{\noindent #1\vspace{5mm}

}
\renewcommand{\abstract}[1]{\vspace{19mm}

\noindent{\small{\em Abstract.} #1}\vspace{2mm}

} 

\def\case#1#2{{\textstyle{#1\over #2}}}

\begin{document}
%
%
\title{Parasupersymmetric Quantum Mechanics\\[2mm]
          with Generalized Deformed Parafermions}
\authors{Jules Beckers \footnote{E-mail: beckers@vm1.ulg.ac.be}, 
Nathalie Debergh \footnote{Chercheur, Institut Interuniversitaire des
Sciences Nucl\'eaires}}
\address{Theoretical and Mathematical Physics, Institute of Physics (B.5),
University of Li\`ege, B-4000 Li\`ege 1, Belgium}
\authors{and Christiane Quesne \footnote{Directeur de recherches, Fonds National de
la Recherche Scientifique; E-mail: cquesne@ulb.ac.be}}
\address{Physique Nucl\'eaire Th\'eorique et Physique Math\'ematique, 
Universit\'e Libre de Bruxelles, Campus de la Plaine CP229, Boulevard du 
Triomphe, B-1050 Brussels, Belgium}
%
%
\abstract{ 
A superposition of bosons and generalized deformed parafermions corresponding
to an arbitrary paraquantization order $p$ is considered to provide
deformations of parasupersymmetric quantum mechanics. New families of
parasupersymmetric Hamiltonians are constructed in connection with two
examples of su(2) nonlinear deformations such as introduced by Polychronakos
and Ro\v cek.  
}
%
%
\section{Introduction}
During the last few years, nonlinear deformations of (the universal enveloping
algebra of) Lie algebras have attracted a lot of attention (see Ref.~\cite{1},
and references therein). They include some specific deformations with a Hopf
algebraic structure, often called $q$-algebras and related to quantum
groups~\cite{2}, as well as more general deformations, such as those of su(2)
introduced by Polychronakos~\cite{3} and Ro\v cek~\cite{4}.\par
%
%
Biedenharn~\cite{5} and Macfarlane~\cite{6} pioneering works on the
$q$-deformed harmonic oscillator have been extended by various authors. Besides
a specific attempt~\cite{7}, the introduction of the generalized deformed
oscillator~\cite{8} has proved useful to provide a unified description~\cite{9}
of the Bose, Fermi, parabose and parafermi harmonic oscillators~\cite{10}, as
well as their $q$-deformations~\cite{11}.\par
%
%
Quite recently~\cite{12}, generalized deformed parafermions, including the
$q$-deformed ones as a special case, were defined in the generalized deformed
oscillator framework, and were shown to be related to some unitary irreducible
representations (unirreps) of the Polychronakos and Ro\v cek su(2)
deformations. Moreover, some physically relevant exactly solvable Hamiltonians,
such as the Morse and modified P\" oschl-Teller ones, were proved to be
equivalent to Fermi-like oscillator Hamiltonians constructed in terms of these
generalized deformed parafermions, which therefore provide a new algebraic
description of their bound state spectrum~\cite{12}.\par
%
%
In the present paper, we will consider a superposition of standard bosons and
such generalized deformed parafermions of arbitrary paraquantization order~$p$
to study deformations of parasupersymmetric quantum mechanics, as initiated in
a previous work~\cite{13}. Our approach will differ with respect to that using
bosons and $q$-deformed parafermions of order~$p$, developed in a recent
paper~\cite{14}.\par
%
%
Our purpose will be twofold: firstly, to introduce generalized deformations of
the parasuperalgebra Psqm(2), generated by two (parasuper)charges $Q$ and
$Q^{\dagger}$, and a parasupersymmetric Hamiltonian $H$, and to establish a
general (necessary and sufficient) condition for the existence of new
(nontrivial) types of Hamiltonians; secondly, to work out in some detail two
examples, connected with two Polychronakos and Ro\v cek
algebras~\cite{3},~\cite{4} already considered before~\cite{12}, in order to
point out how some new (nontrivial) properties and results can be obtained.\par
%
%
\section{Generalized Deformations of the Parasuperalgebra Psqm(2)}
Let $b$ and $b^{\dagger}$ denote generalized deformed parafermionic operators,
as defined in Proposition~2 of Ref.~\cite{12}, i.e., operators satisfying the
nilpotency relations
\begin{equation}
  b^{p+1} = 0 \qquad \bigl(b^{\dagger}\bigr)^{p+1} = 0 \label{eq:1}
\end{equation}
and the trilinear relations
\begin{equation}
  \bigl[b, \bigl[b^{\dagger},b\bigr] \bigr] = G(N) b \qquad
  \bigl[b^{\dagger}, \bigl[b,b^{\dagger}\bigr] \bigr] = b^{\dagger} G(N) 
  \label{eq:2}
\end{equation}
where
\begin{equation}
  G(N) = 2F(N+1) - F(N) - F(N+2) \qquad F(N) = b^{\dagger} b \label{eq:3}
\end{equation}
$F$ being any positive analytic function \footnote{In Ref.~\cite{12}, $F$
was assumed to be strictly positive on the set $\{1,2,\ldots,p\}$. Here, we
shall only assume that it is nonnegative.}.\par
%
%
Let us recall that for standard parafermions
\begin{equation}
  G(N) = 2 \qquad F(N) = F(p+1-N) = N (p+1-N) \label{eq:4}
\end{equation}
so that relations~(\ref{eq:1})--(\ref{eq:3}) reduce to the original
characterization of parastatistics~\cite{10}. In general, we shall replace
the factor $k(p+1-k)$ by the function $F(p+1-k)$, so that the matrix
realization of the generalized deformed parafermionic operators (see
Ref.~\cite{14}) is given by
\begin{equation}
  b = \sum_{k=1}^p \bigl[F(p+1-k)\bigr]^{1/2} e_{k+1,k} \qquad
  b^{\dagger} = \sum_{k=1}^p \bigl[F(p+1-k)\bigr]^{1/2} e_{k,k+1} \label{eq:5}
\end{equation}
in terms of $(p+1)$-dimensional matrices $e_{m,n}$ with entry~1 at the
intersection of row~$m$ and column~$n$ and zeroes everywhere else.\par
%
%
Let us modify the generalized parasupercharges $Q$ and $Q^{\dagger}$ introduced
in the $q$-deform\-ations of Psqm(2) study~\cite{14}, by substituting the
operators~(\ref{eq:5}) for the $q$-deformed para\-fermionic operators. Hence, $Q$
and $Q^{\dagger}$ are now written as
\begin{eqnarray}
  Q & = & \sum_{k=1}^p \left[\case{1}{2}F(p+1-k)\right]^{1/2} \bigl(p_x + i
      W_k(x)\bigr)\, e_{k+1,k} \nonumber \\
  Q^{\dagger} & = & \sum_{k=1}^p \left[\case{1}{2}F(p+1-k)\right]^{1/2} \bigl(p_x
      - i W_k(x)\bigr)\, e_{k,k+1} \label{eq:6}
\end{eqnarray}
where $p_x = -i d/dx$, and $W_k(x)$, $k=1$, 2, $\ldots$,~$p$, refer to the
parasuperpotentials inside the bosonic operators. In correspondence with
eqs.~(\ref{eq:1})--(\ref{eq:3}), such charges have to satisfy the nilpotency
relations
\begin{equation}
  Q^{p+1} = 0 \qquad \bigl(Q^{\dagger}\bigr)^{p+1} = 0 \label{eq:7}
\end{equation}
as well as the structure ones
\begin{equation}
  \bigl[Q, \bigl[Q^{\dagger},Q\bigr] \bigr] = G(N) Q H \qquad
  \bigl[Q^{\dagger}, \bigl[Q,Q^{\dagger}\bigr] \bigr] = Q^{\dagger} H G(N) 
  \label{eq:8}
\end{equation}
where $H$ plays the role of the deformed parasupersymmetric Hamiltonian, with
respect to which the charges are conserved, i.e.,
\begin{equation}
  \bigl[H, Q\bigr] = 0 \qquad \bigl[H, Q^{\dagger}\bigr] = 0. \label{eq:9}
\end{equation}
Let us point out that relations (\ref{eq:7})--(\ref{eq:9}) characterize the
deformed parasupersymmetric algebra associated with our superposition of bosons
and generalized deformed parafermions.\par
%
%
As in the $q$-deformed case~\cite{14}, the structure relations~(\ref{eq:8})
imply some constraints on the superpotentials in the form of Riccati equations
\begin{equation}
  W_{k+1}^2 + W'_{k+1} = W_k^2 - W'_k + c_k \qquad k = 1,2,\ldots,p-1
               \label{eq:10}
\end{equation}
where primes refer to space derivatives, and the $c_k$'s are so far arbitrary
constants. The latter will be chosen by requiring that the parasupersymmetric
Hamiltonian is diagonal with non-vanishing matrix elements given by
\begin{equation}
  H_{k,k} = \case{1}{2} p_x^2 + f_k(x) \qquad k = 1,2,\ldots,p+1
           \label{eq:11}
\end{equation}
in terms of some functions $f_k(x)$ to be determined.\par
%
%
By introducing eq.~(\ref{eq:6}) into the two trilinear relations~(\ref{eq:8}),
we obtain after some relatively tedious calculations two sets of $p$ equations,
given by
\begin{eqnarray}
  G(p-k) H_{k,k} & = & \case{1}{2} G(p-k) \bigl(p_x^2 + W_k^2 + W'_k\bigr)
     + \case{1}{2} \bigl[F(p+2-k) c_{k-1} - F(p-k) c_k\bigr] \nonumber \\
  &  & k = 1,2,\ldots,p \label{eq:12}
\end{eqnarray}
and
\begin{eqnarray}
  G(p+1-k) H_{k,k} & = & \case{1}{2} G(p+1-k) \bigl(p_x^2 + W_k^2 + W'_k\bigr)
     + \case{1}{2} \bigl[F(p+3-k) (c_{k-2} - c_{k-1}) \nonumber \\
  & & - 2 F(p+2-k) c_{k-1}\bigr] \qquad k = 2,3,\ldots,p \nonumber \\
  G(0) H_{p+1,p+1} & = & \case{1}{2} G(0) \bigl(p_x^2 + W_p^2 - W'_p\bigr)
     + \case{1}{2} F(2) c_{p-1} \label{eq:13}
\end{eqnarray}
respectively. The latter lead to two different expressions for $H_{k,k}$,
$k=2$, 3, $\ldots$,~$p$. By equating them, we get a system of $p-1$ homogeneous
linear equations in $p-1$ unknowns~$c_k$, $k=1$, 2, $\ldots$,~$p-1$.\par
%
%
Such a system always admits the trivial solution $c_k=0$, $k=1$, 2,
$\ldots$,~$p-1$. For the corresponding parasupersymmetric Hamiltonian, we then
recover the original result of undeformed Psqm(2) for an arbitrary
paraquantization order~\cite{14}. The system however also admits a nontrivial
solution (i.e., with at least one nonzero arbitrary constant), absent in the
undeformed case, if and only if the determinant of its coefficients vanishes.
We did establish the general form of this necessary and sufficient condition.
For brevity's sake, we only quote here the final result:
\begin{equation}
  (p+1) F(1) F(2) \ldots F(p) G(1) G(2) \ldots G(p-2) = 0. \label{eq:14}
\end{equation}
We conclude that a nontrivial solution does exist if and only if one of the
arbitrary functions $F(k)$ or $G(k)$ vanishes for some $k\in \{1,2,\ldots,p\}$ or
$\{1,2,\ldots,p-2\}$, respectively. The diagonal elements of the corresponding
parasupersymmetric Hamiltonian are then given by either equation~(\ref{eq:12})
or~(\ref{eq:13}).\par
%
%
In the next section, we will show on two examples that nontrivial solutions do
indeed exist.\par
%
%
\section{Examples}
The examples to be considered here correspond to generalized
deformed parafermionic operators transforming under a ($p+1$)-dimensional
unirrep of some Polychronakos~\cite{3} and Ro\v cek~\cite{4} deformed su(2)
algebra (instead of su(2), as in the undeformed case). Such a nonlinear algebra
is defined by the commutation relations
\begin{equation}
  [J_0, J_{\pm}] = \pm J_{\pm} \qquad [J_+, J_-] = f(J_0) \label{eq:15}
\end{equation}
where $f(J_0)$ is some real, analytic function in~$J_0$, going to $2J_0$ for
some limiting values of the parameters. Whenever $f(J_0)$ is a polynomial of
degree less than or equal to three, the functions $F(N)$ and $G(N)$ of
eqs.~(\ref{eq:2}) and~(\ref{eq:3}), characterizing the generalized deformed
parafermions, are given by \footnote{In eq.~(\ref{eq:16}), some misprints in
Ref.~\cite{12} have been corrected.}
\begin{eqnarray}
  F(N) & = & N (p+1-N) (\lambda + \mu N + \nu N^2) \nonumber \\
  G(N) & = & 2 \left\{\lambda - (p-2) \mu - (3p-4) \nu + 3 \bigl[\mu - (p-3) \nu
     \bigr] N + 6 \nu N^2\right\} \label{eq:16}
\end{eqnarray}
where the constants $\lambda$, $\mu$,~$\nu$ can be found from $f(J_0)$ as
explained in Proposition~5 of Ref.~\cite{12}.\par
%
%
The first example \cite{4}, \cite{12} corresponds to
\begin{equation}
  f(J_0) = 2J_0 + \alpha J_0^2 \label{eq:17}
\end{equation}
where
\begin{equation}
  |\alpha| < {6\over 2p+1} \label{eq:18}
\end{equation}
ensures the existence of a ($p+1$)-dimensional unirrep characterized by a
highest weight $j = {1 \over2}p - \alpha^{-1}(1-\epsilon)$, where $\epsilon =
\left[1 - {1\over12}\alpha^2 p(p+2)\right]^{1/2}$. The functions $F(N)$ and
$G(N)$ are then given by~(\ref{eq:16}), where $\lambda = - {1\over6}\alpha
(p+1) + \epsilon$, $\mu = {1\over3}\alpha$, and~$\nu=0$.\par
%
%
As $F(k)>0$ for $k=1$, 2, $\ldots$,~$p$, since the representation considered is
irreducible and unitary, condition~(\ref{eq:14}) can only be satisfied provided
$G(k)$ vanishes, i.e.,
\begin{equation}
  \epsilon = \case{1}{2} (p-1-2k)\,\alpha \label{eq:19}
\end{equation}
for some $k\in \{1,2,\ldots,p-2\}$. Condition~(\ref{eq:19}) is equivalent to
\begin{equation}
  \alpha = 2 \sqrt{3}\,\sigma \left[4p^2 - 4(3k+1)p + 3(2k+1)^2\right]^{-1/2}
          \label{eq:20}
\end{equation}
where $\sigma$ denotes the sign of ${1\over2}(p-1)-k$. The
inequality~(\ref{eq:18}) here requires
\begin{equation}
  (2p-3k-2)(p-3k-1) > 0 \label{eq:21}
\end{equation}
which is possible only for $p\ge5$. For an arbitrary paraquantization order
$p=3l-1$, $3l$, or~$3l+1$, where~$l\ge2$, the allowed parameter values leading
to new parasupersymmetric Hamiltonians are therefore given by~(\ref{eq:20}),
where $k=1$, 2, $\ldots$,~$l-1$ or $k=p-l$, $p-l+1$, $\ldots$,~$p-2$.\par
%
%
The minimal context $p=5$ only allows the values $k=1$ and $k=3$, leading to
specific parameter values $\alpha = 2 \sqrt{3/47}$, and $\alpha = -2 \sqrt{3/47}
$, respectively. In such a case, we therefore obtain two new families of
parasupersymmetric Hamiltonians, whose diagonal matrix elements are given by
\begin{eqnarray}
  H_{i,i} & = & \case{1}{2} \bigl(p_x^2 + W_i^2 + W'_i\bigr) + \gamma_i \qquad 
                   i=1,2,\ldots, 5 \nonumber \\
  H_{6,6} & = & \case{1}{2} \bigl(p_x^2 + W_5^2 - W'_5\bigr) + \gamma_5 
                   \label{eq:22}
\end{eqnarray}
where 
\begin{equation}
  \gamma_1 = -\case{8}{9} c_1 \qquad \gamma_2 = -\case{25}{18} c_1 \qquad
       \gamma_3 = -\case{200}{81} c_1 \qquad \gamma_4 = -\case{50}{9} c_1 \qquad
       \gamma_5 = -\case{400}{9} c_1 \label{eq:23}
\end{equation}
and
\begin{equation}
  \gamma_1 = \case{4}{3} c_1 \qquad \gamma_2 = \case{5}{6} c_1 \qquad
       \gamma_3 = \case{2}{27} c_1 \qquad \gamma_4 = \case{1}{24} c_1 \qquad
       \gamma_5 = \case{2}{75} c_1 \label{eq:24}
\end{equation}
respectively.\par
%
%
Let us also notice that besides these general results, two other specific cases 
have to be distinguished. They are associated with the possible vanishing of
either $G(0) = 2F(1) - F(0)$ or $G(p-1) = 2F(p) - F(p-1)$, and correspond to the
parameter values $\alpha = 2 \sqrt{3} (4p^2-4p+3)^{-1/2}$, or $\alpha = -2
\sqrt{3} (4p^2-4p+3)^{-1/2}$. Such values satisfy the inequality~(\ref{eq:18}),
but leave the matrix element $H_{p+1,p+1}$ or $H_{1,1}$ entirely arbitrary,
thereby excluding the knowledge of the corresponding parasupersymmetric
Hamiltonian final form.\par 
%
%
The second example~\cite{12} corresponds to
\begin{equation}
  f(J_0) = 2J_0 + \alpha J_0^3 \label{eq:25}
\end{equation}
where
\begin{equation}
  \alpha > -{8\over p^2} \label{eq:26}
\end{equation}
ensures the existence of a ($p+1$)-dimensional unirrep characterized by $j=p/2$.
The corresponding functions $F(N)$ and $G(N)$ are given by~(\ref{eq:16}), where
$\lambda = 1 + \case{1}{8} \alpha p(p+2)$, $\mu = -\case{1}{4} \alpha (p+1)$, 
and $\nu = \case{1}{4} \alpha$.\par
%
%
Once again, condition~(\ref{eq:14}) will be satisfied provided $G(k)=0$, i.e.,
\begin{equation}
  \alpha = -8 \bigl[3p^2 - 6(2k+1)p + 12k^2 + 12k + 4]^{-1} \label{eq:27}
\end{equation}
for some $k \in \{1,2,\ldots,p-2\}$. Such a parameter value satisfies the
inequality~(\ref{eq:26}) if
\begin{equation}
  p < \case{3}{2} (2k+1) - \case{1}{2} (12k^2 + 12k + 1)^{1/2} \qquad 
        \mbox{\rm or}
        \qquad p > \case{3}{2} (2k+1) + \case{1}{2} (12k^2 + 12k + 1)^{1/2}.
        \label{eq:28}
\end{equation}
A detailed discussion of these conditions leads to the result that the allowed
parameter values giving rise to new parasupersymmetric Hamiltonians correspond
to $k=1$, 2, $\ldots$,~$l-1$, or $k=p-l$, $p-l+1$, $\ldots$,~$p-2$, and
\begin{equation}
  4l + \kappa_l \le p < 4(l+1) + \kappa_{l+1} \label{eq:29}
\end{equation}
where $\kappa_l$ is the integer determined by the condition
\begin{equation}
  - l - \case{3}{2} + \case{1}{2}(12l^2 - 12l + 1)^{1/2} < \kappa_l \le - l - 
         \case{1}{2} + \case{1}{2}(12l^2 - 12l + 1)^{1/2}. \label{eq:30}
\end{equation} 
We also observe that the matrix elements $H_{1,1}$ and $H_{p+1,p+1}$ are left
arbitrary if $p>2$ and $\alpha = -8 (3p^2 - 6p + 4)^{-1}$.\par
%
%
The lowest paraquantization order for which conditions~(\ref{eq:29})
and~(\ref{eq:30}) are satisfied is $p=8$. Then $G(1) = G(6) = 0$ for $\alpha = 
-2/19$. The corresponding family of new parasupersymmetric Hamiltonians is
defined by the following set of diagonal matrix elements:
\begin{eqnarray}
  H_{i,i} & = & \case{1}{2} \bigl(p_x^2 + W_i^2 + W'_i\bigr) + \gamma_i \qquad
                   i=1,2,\ldots, 8 \nonumber \\
  H_{9,9} & = & \case{1}{2} \bigl(p_x^2 + W_8^2 - W'_8\bigr) + \gamma_8 
                   \label{eq:31}
\end{eqnarray}
where 
\begin{eqnarray}
  \gamma_1 & = & \case{7}{6} c_1 \qquad \gamma_2 = \case{2}{3} c_1 \qquad
       \gamma_3 = \case{7}{24} c_1 - \case{15}{4} c_3 \qquad \gamma_4 =
       \case{7}{24} c_1 - \case{17}{4} c_3 \nonumber \\
  \gamma_5 & = & \case{7}{20} c_1 - \case{11}{2} c_3 \qquad \gamma_6 =
       \case{49}{96} c_1 - \case{135}{16} c_3 \qquad \gamma_7 = c_1
       - \case{120}{7} c_3 \nonumber \\
  \gamma_8 & = & \case{329}{48} c_1 - \case{975}{8} c_3. \label{eq:32} 
\end{eqnarray}
\par
%
%
\section{Some Comments}
In conclusion, we did show through two examples that the approach developed in
the present paper leads to families of new deformed parasupersymmetric
Hamiltonians. As compared with those generated by the previous approach based
upon $q$-deformed parafermions \cite{14}, the latter correspond to rather high
paraquantization orders.\par
%
%
As a last comment, we would like to mention that the $p=2$ case can be discussed
in full details. The corresponding system obtained from~(\ref{eq:12})
and~(\ref{eq:13}), as well as condition~(\ref{eq:14}) lead to the following three
Hamiltonians
\begin{equation}
  H^{(1)} = \case{1}{2} p_x^2 + \case{1}{2}\left(
    \begin{array}{ccc}
        W_1^2 + W'_1 & 0                              & 0                             \\
        0                    & W_2^2 + W'_2 - c_1 & 0                              \\
        0                    & 0                              & W_2^2 - W'_2 - c_1
    \end{array}
  \right) \label{eq:33}
\end{equation}
if $F(1) = 0$,
\begin{equation}
  H^{(2)} = \case{1}{2} p_x^2 + \case{1}{2}\left(
    \begin{array}{ccc}
        W_1^2 + W'_1 + c_1 & 0                    & 0                             \\
        0                             & W_2^2 + W'_2 & 0                              \\
        0                             & 0                    & W_2^2 - W'_2 
    \end{array}
  \right) \label{eq:34}
\end{equation}
if $F(2) = 0$, and
\begin{equation}
  H^{(3)} = \case{1}{2} p_x^2 + \case{1}{2}\left(
    \begin{array}{ccc}
        W_1^2 + W'_1 & 0                    & 0                             \\
        0                    & W_2^2 + W'_2 & 0                              \\
        0                    & 0                    & W_2^2 - W'_2 
    \end{array}
  \right) \label{eq:35}
\end{equation}
otherwise. Let us notice that the first two cases lead to pseudostatistical
considerations~\cite{15}. 
\par
%
%
For oscillator-like interactions characterized by $W_1 = \omega x$ ($\omega$ being
the angular frequency), the Hamiltonians (\ref{eq:33})--(\ref{eq:35}) become
\begin{eqnarray}
  H^{(1)} & = & \case{1}{2} p_x^2 + \case{1}{2} \omega^2 x^2 + \case{1}{2} \omega
      \left(\begin{array}{rrr}
          1 & 0   & 0    \\
          0 & -1 & 0     \\
          0 & 0   & -3
      \end{array}\right) \label{eq:36} \\
  H^{(2)} & = & \case{1}{2} p_x^2 + \case{1}{2} \omega^2 x^2 + \case{1}{2} \omega
      \left(\begin{array}{rrr}
          3 & 0 & 0    \\
          0 & 1 & 0     \\
          0 & 0 & -1
      \end{array}\right) \label{eq:37} \\
  H^{(3)} & = & \case{1}{2} p_x^2 + \case{1}{2} \omega^2 x^2 + \case{1}{2} \omega
      \left(\begin{array}{rrr}
          1 & 0   & 0    \\
          0 & -1 & 0     \\
          0 & 0   & 1
      \end{array}\right) \label{eq:38} 
\end{eqnarray}
respectively. They can be interpreted as Hamiltonians of a system consisting of a
non-interacting three-level subsystem and one bosonic mode, as
occurring in quantum optics~\cite{16}. The third possibility~(\ref{eq:38})
corresponds to the so-called ${}\vee{}$-type and is the only one available in the
undeformed and $q$-deformed~\cite{5,6} contexts. Our general
deformation~(\ref{eq:6}) completes the information by allowing the other possible
scheme of three-level configurations, namely the $\Xi$-type.\par
%
%
Moreover, the Hamiltonian~$H^{(1)}$ supplemented by the constant term
$\case{1}{2} \omega$ (or, equivalently, $H^{(2)}$ supplemented by $-\case{1}{2}
\omega$) has a clear physical interpretation as describing the motion of a spin-1
particle in both an oscillator potential and a homogeneous magnetic
field~\cite{17}. Once again, this result is relevant to our general
deformation~(\ref{eq:6}), but is not possible either in the undeformed or the
$q$-deformed context.\par
%
%
Whether some other examples, associated with specific sets of analytic functions
$F(N)$, may be of physical interest remains an open question, to which we hope to
come back in a near future.\par
\newpage
%
%


\begin{thebibliography}{99}

\bibitem{1} V. Chari and A. Pressley, A guide to quantum groups (Cambridge U.P.,
Cambridge, 1994).

\bibitem{2} V. G. Drinfeld, in: Proc. Int. Cong. of Mathematicians (Berkeley, CA, 1986),
ed. A. M. Gleason (AMS, Providence, RI, 1987) p. 798; \\
M. Jimbo, Lett. Math. Phys. {\bf 10} (1985) 63; {\bf 11} (1986) 247.

\bibitem{3} A. P. Polychronakos, Mod. Phys. Lett. A {\bf 5} (1990) 2325.

\bibitem{4} M. Ro\v cek, Phys. Lett. B {\bf 255} (1991) 554.

\bibitem{5} L. C. Biedenharn, J. Phys. A {\bf 22} (1989) L873.

\bibitem{6} A. J. Macfarlane, J. Phys. A {\bf 22} (1989) 4581.

\bibitem{7} J. Beckers and N. Debergh, J. Phys. A {\bf 24} (1991) L1277.

\bibitem{8} C. Daskaloyannis, J. Phys. A {\bf 24} (1991) L789.

\bibitem{9} D. Bonatsos and C. Daskaloyannis, Phys. Lett. B {\bf 307} (1993) 100.

\bibitem{10} H. S. Green, Phys. Rev. {\bf 90} (1953) 270; \\
Y. Ohnuki and S. Kamefuchi, Quantum field theory and parastatistics (Springer,
Berlin, 1982).

\bibitem{11} R. Floreanini and L. Vinet, J. Phys. A {\bf 23} (1990) L1019; \\
E. Celeghini, T. D. Palev and M. Tarlini, Mod. Phys. Lett. B {\bf 5} (1991) 187; \\
K. Odaka, T. Kishi and S. Kamefuchi, J. Phys. A {\bf 24} (1991) L591; \\
M. Krishna Kumari, P. Shanta, S. Chaturvedi and V. Srinivasan, Mod. Phys. Lett. A
{\bf 7} (1992) 2593.

\bibitem{12} C. Quesne, Phys. Lett. A {\bf 193} (1994) 245.

\bibitem{13} J. Beckers and N. Debergh, Nucl. Phys. B {\bf 340} (1990) 767.

\bibitem{14} J. Beckers and N. Debergh, J. Phys. A {\bf 26} (1993) 4311.

\bibitem{15} J. Beckers and N. Debergh, Int. J. Mod. Phys. A {\bf 10} (1995) 2783; \\
J. Beckers, N. Debergh and A. G. Nikitin, Fortschr. Phys. {\bf 43} (1995) 81.

\bibitem{16} H.-I. Yoo and J. H. Eberly, Phys. Rep. {\bf 118} (1985) 239; \\
V. V. Semenov and S. M. Chumakov, Phys. Lett. B {\bf 262} (1991) 451.

\bibitem{17} V. A. Rubakov and V. P. Spiridonov, Mod. Phys. Lett. A {\bf 3} (1988)
1337.

\end {thebibliography}
\end{document}